\documentclass {article}

\usepackage[latin1]{inputenc}
\usepackage[T1]{fontenc}
\usepackage{amsmath}
\usepackage{amsfonts}
\usepackage{amssymb}
\usepackage{latexsym}
\usepackage[english]{babel}
\usepackage[dvips]{graphicx,color}
\usepackage{latexsym}
\usepackage{graphicx}
\usepackage{graphics}
\usepackage{pstricks}

\newcommand{\be}{\begin{equation}}
\newcommand{\ee}{\end{equation}}
\newcommand{\beq}{\begin{equation}}
\newcommand{\eeq}{\end{equation}}
\newcommand{\bea}{\begin{eqnarray}}
\newcommand{\eea}{\end{eqnarray}}
\newcommand{\nn}{\nonumber}
\newcommand{\ba}{\begin{eqnarray}}
\newcommand{\ea}{\end{eqnarray}}
\def\bc{\begin{center}}
\def\ec{\end{center}}

\begin{document}

\begin{titlepage}

\vspace{2in}

\begin{centering}

{\Large {\bf Lorentzian AdS, Wormholes and Holography}}

\vspace{.3in}

Ra{\'u}l E. Arias $^{\dag}$, Marcelo Botta Cantcheff $^{\ddag\,\dag}$ and Guillermo A. Silva $^{\dag}$\\

\vspace{.2 in}
$^{\dag}${\it IFLP-CONICET and Departamento de F\'{\i}sica\\ Facultad de Ciencias Exactas, Universidad Nacional de La Plata\\
CC 67, 1900,  La Plata, Argentina}\\
$^{\ddag}${\it CERN, Theory Division, 1211 Geneva 23, Switzerland}
\vspace{.4in}
%


\begin{abstract}

We investigate the structure of two point functions for the QFT dual
to an asymptotically Lorentzian AdS-wormhole. The bulk geometry is a
solution of 5-dimensional second order Einstein Gauss Bonnet gravity
and causally connects two asymptotically AdS space times. We revisit
the GKPW prescription for computing two-point correlation functions
for dual QFT operators $\cal O$ in Lorentzian signature and we
propose to express the bulk fields in terms of the independent
boundary values $\phi_0^\pm$ at each of the two asymptotic AdS
regions, along the way we exhibit how the ambiguity of normalizable
modes in the bulk, related to initial and final states, show up in
the computations. The independent boundary values are interpreted as
sources for dual operators $\cal O^\pm$ and we argue that, apart
from the possibility of entanglement, there exists a coupling
between the degrees of freedom leaving at each boundary.
 The AdS$_{1+1}$ geometry
is also discussed in view of its similar boundary structure.
Based on the analysis, we propose a very simple geometric criterium
to distinguish coupling from entanglement effects among the two set of degrees
of freedom associated to each of  the disconnected parts of the boundary.

\end{abstract}

\end{centering}

~

\noindent

\end{titlepage}
\section{Introduction}

Asymptotically AdS geometries play an important role in the
gauge/gravity correspondence \cite{adscft,GKP,eddie}, since they
provide gravity duals to quantum field theories (QFT) with  UV
conformal fixed points. There is a general consensus, based on several
checks, for the dual interpretation of various asymptotically AdS geometries: a big black
hole solution is supposed to describe a thermal QFT state \cite{wittenT},
a bulk solution interpolating between an AdS horizon (corresponding
to an IR conformal field fixed point) and an AdS geometry at infinity
of different radii realizes the renormalization group flow
between two conformal fixed points \cite{pw}. As a third possibility,
certain regular (solitonic)  charged AdS solutions are interpreted
as excited QFT coherent states \cite{llm}.

We would like to discuss in this work the more intriguing situation
that appears when a wormhole in the bulk causally connects two
asymptotic AdS$_{d+1}$ (Lorentzian) boundaries. Holography and the
AdS/CFT correspondence in the presence of multiple boundaries is
less understood. The implementation of the AdS/CFT paradigm for such
cases suggests that the dual field theory lives on the union of the
disjoint boundaries, and therefore to be the product of field
theories on the different boundaries (see \cite{deboer}). We would
revisit this statement  and discuss the issue of whether the two
dual theories are independent, decoupled or not.

For Lorentzian signature, wormhole geometries  are ruled out for
$d\geq2$ dimensions as a solution of an Einstein-Hilbert action
satisfying natural causality conditions: disconnected boundaries
must be separated by horizons \cite{galloway} (see \cite{svr} for
recent work in 2+1 dimensions). Studies of wormholes in string
theory and in the context of the AdS/CFT correspondence have
therefore concentrated on Euclidean signature spaces particulary
motivated from applications to cosmology (see references to
\cite{maldamaoz,polchi}). For completeness we quote that in the
Euclidean context a theorem states that disconnected positive scalar
curvature boundaries are also ruled for complete Einstein manifolds
of negative curvature \cite{WittenYau} (see also \cite{Cai}).
Moreover, \cite{WittenYau} proves that for negative curvature
boundaries the holographic theory living on them would be unstable
for $d\geq3$ (see also \cite{buchel}). The wormholes studied in
\cite{maldamaoz} avoided the theorem in \cite{WittenYau} since they
were constructed as hyperbolic slicings of AdS and supported by
extra supegravity fields.

The canonical Lorentzian example of two boundaries separated by a
horizon is the Eternal black hole geometry and the proposal
put forward in \cite{Eternal} makes contact with the
thermo field dynamics (TFD) formulation of QFT at finite
temperature \cite{TFD}: the two disconnected boundaries amount to
two decoupled copies ${\cal H}_{\pm}$ of the dual field theory and non
vanishing correlators  $\langle {\cal O}^+({\mathbf x})\, {\cal O}^-({\mathbf x'})\rangle$ are
interpreted as being averaged over an entangled state encoding
the statistical/thermal information of the bulk geometry (see also
\cite{Azeyanagi} and \cite{vaman} for recent work).
An interesting second Lorentzian
example with two disconnected boundaries was constructed in
\cite{balasu} by performing a non-singular orbifold of AdS$_3$. The
result of the construction led to two causally connected
cylindrical boundaries with the dual field theory involving the DLCQ
limit of the D1-D5 conformal field theory,  but the coupling
 between the different boundaries degrees of freedom was not clarified. The main difference between these
two examples is that in the last case causal contact exists between the conformal boundaries.
The no-go theorem \cite{galloway} is bypassed in the second case because the performed
quotient results in the presence of compact direction with the
geometry effectively being a $S^1$ fibration over AdS$_2$ where the
aforementioned theorem does not apply.

The no-go Lorentzian wormholes theorem \cite{galloway} is also bypassed when
working with a higher order gravity theory, moreover,
higher order curvature corrections to standard Einstein gravity are generically expected for
any quantum theory of gravity.
However, not much is known about the precise forms of the higher derivative
corrections, other than for a few maximally supersymmetric cases. Since from the pure
gravity point of view the most general theory that leads to second-order field equations for the metric is
of the Lovelock type \cite{love}, we will choose to work with
the simplest among them known as Einstein-Gauss-Bonnet theory.
The action for this theory only contains terms up to quadratic order in the curvature and
our interest in the wormhole solution, found in \cite{Dotti}, is that its simplicity
permits an analytic treatment. The geometry corresponds to a static wormhole
connecting two asymptotically AdS regions with base manifold $\tilde\Sigma$, which in $d+1=5$ takes the form
$\tilde\Sigma= H^3$ or $S^1\times H^2$, where
$H^2$ and $H^3$ are two and three-dimensional (quotiented) hyperbolic spaces. The resulting geometry is smooth,
does not contain horizons anywhere, and the two asymptotic regions turn out to be causally
connected. A perturbative stability study for the 5-dimensional solution case of \cite{Dotti}
was performed in \cite{diego}.

We will revisit in the present paper the
Gubser-Klebanov-Polyakov-Witten (GKPW) prescription \cite{GKP,eddie}
for extracting QFT correlators from gravity computations and discuss
its application for the Lorentzian wormhole solution  found in
\cite{Dotti} mentioning along the way the similarities and
differences  with the AdS$_2$ case (see \cite{diego,ali} for other
work on the wormhole background discussed here). We recall that the
GKPW prescription in Lorentzian signature involves not only boundary
data at the conformal boundary of the spacetime but also the
specification of initial and final states, we will show how these
states  make their appearance in the computations (see
\cite{Herzog,bala,bala2,balagid,marolf,Skenderis} for discussions on
Lorentzian issues related to the GKWP prescription). It is commonly
accepted that the QFT dual to a wormhole geometry should correspond
to two independent gauge theories living at each boundary and the
wormhole geometry encodes an entangled state among them. On the
other hand, the causal connection between the boundaries has been
argued to give rise to a non-trivial coupling between the two
dual theories \cite{balasu}. We will argue, by performing
an analytic continuation to the Euclidean section of the space
time, that the non vanishing result obtained for the
correlator $\langle {\cal O}^+({\mathbf x})\, {\cal O}^-({\mathbf x'})\rangle$ between
operators located at opposite boundaries signals the existence of a
coupling between the fields associated to each boundary.

The paper is organized as follows: in section \ref{GKPW} we review
the GKPW prescription for computing QFT correlation functions from
gravity computations mentioning the peculiarities of Lorentzian
signature, in section \ref{gkpww} we extend the GKPW prescription
for the case of two asymptotic independent boundary data. We apply
it to $AdS_2$, reproducing the results appearing in the literature,
and to the wormhole \cite{Dotti} showing their similarities. In
section \ref{holentanglement} we discuss several arguments regarding
the possibility of entanglement and/or interactions among the two
dual QFT. We summarize in section \ref{discussion} the results of the paper.

\section{GKPW prescription with a single asymptotic boundary}
\label{GKPW} The GKPW prescription \cite{GKP,eddie} instructs to
equates the gravity (bulk) partition function for an asymptotically
AdS$_{d+1}$ spacetime ${\cal M}$, understood as a functional of
boundary data, to the generating functional for correlators of a CFT
defined on the spacetime conformal boundary $\partial\cal M$.
Explicitly the prescription is \be
\mathcal{Z}_{gravity}\left[\phi(\phi_0)\right]=\langle
e^{i\int_{\partial\cal
M}d^d{\mathbf{x}}\;\phi_0(\mathbf{x})\mathcal{O(\mathbf{x})}}\rangle\;.
\label{partition} \ee In the left hand side
$\phi_0=\phi_0(\mathbf{x})$ stands for the boundary value of the
field $\phi$, and the right hand side is the CFT generating
functional of correlators  of the operator $\mathcal{O}$ dual to the
(bulk) field $\phi$. In the present paper we will be working in the
semiclassical spacetime limit (large N limit of the CFT) and
therefore the lhs in \eqref{partition} will be aproximated by the
on-shell action of the field $\phi$ which for simplicity will be
taken to be a scalar field of mass $m$. We are interested in real
time (Lorentzian) geometries and in this case, the prescription
\eqref{partition} is incomplete, since a specification of the
initial and final states $\psi_{\mathrm {i,f}}$ on which we compute
the correlator on the rhs need to be specified, we will discuss this
issue below.

To set out the notation we summarize the prescription for massive scalar fields highlighting the
points important for our arguments. The $AdS_{d+1}$ metric in
Poincar{\'e} coordinates reads
\be
 ds^2=\frac{R^2}{z^2}(d\mathbf{x}^2+ dz^2)\,,
 \label{adsp}
\ee
where the term $d\mathbf{x}^2$ stands for $-dt^2+d\vec{x}^2$. The (conformal) boundary of AdS is
located at $z=0$ and a horizon exists at $z=\infty$\footnote{In the euclidean case $z=\infty$
is just a point, leading to the half plane $z\ge0$ in \eqref{adsp} being compactified to a sphere.}.
The solution to the $\phi$ field equation
subject to boundary data $\phi_0$ set at the conformal boundary is commonly  written as
\be
 \phi(\mathbf{x},z)=\int_{\partial\cal M}d\mathbf{y}\,{\sf K}(\mathbf{x},z\mid\mathbf{y})\,\phi_0(\mathbf{y})\,.
\label{phi}
\ee
In the free field limit, the KG equation shows that the asymptotic behavior for $\phi$ is
\be
\phi(\mathbf{x},z)\sim z^{\Delta_\pm}\phi_0(\mathbf x),~~~~z\to0
\label{asympfi}
\ee
where\footnote{Negative mass scalar fields are allowed in AdS as long as $\mu\ge0$. The minimum allowed mass
for a scalar field in AdS$_{d+1}$ is given by the so called  Breitenlohner-Freedman (BF) bound
$\mu_{BF}=0$, or equivalently $m_{BF}^2=-d^2/4$ \cite{BF}.}
\be
\Delta_\pm=\frac{d}{2}\pm\mu,~~~~~~\mu=\sqrt{\frac{d^2}{4}+m^2R^2}\,.
\label{lambda}
\ee
The bulk-boundary propagator $\sf K$ in \eqref{phi} is therefore demanded to satisfy \cite{GKP,eddie}
\be
(\Box-m^2)\,{\sf K}(\mathbf{x},z\mid \mathbf{y})=0\label{box}
\ee
with boundary condition
\be
 {\sf K}(\mathbf{x},z\mid \mathbf{y})\sim z^{\Delta_-}\delta(\mathbf{x}-\mathbf{y})\,,~~~~z\to0\,.
 \label{K boundary condition}
\ee
Finally, the bulk-boundary propagator $\sf K$ can be related to
the Dirichlet bulk-bulk Green function ${\sf
G}(\mathbf{x},z|\mathbf{y},z')$ through Green's second identity, the
result being that ${\sf K}$ can be obtained from the normal derivative
of ${\sf G}$ evaluated at the spacetime boundary (see
\cite{fdh,mcg}) as
\be
 {\sf K}(\mathbf{x},z\mid \mathbf{y})=\lim_{z'\to0}\sqrt{-g}g^{z'\!z'}\partial_{z'}{\sf
G}\,(\mathbf{x},z\mid\mathbf{y},z')\,.
\label{KfromG}
\ee

Two comments are in order: i) the bulk solution for a given boundary
data $\phi_0$ computed from \eqref{phi} is not unique since
Lorentzian AdS spaces admit normalizable solutions
$\varphi(\mathbf{x},z)$   that can be added at will
to \eqref{phi} without altering the boundary behavior \eqref{K
boundary condition}, explicitly
\be
 \phi(\mathbf{x},z)=\int_{\partial\cal M}d\mathbf{y}\,{\sf K}(\mathbf{x},z\mid\mathbf{y})\,
 \phi_0(\mathbf{y})+\varphi(\mathbf{x},z)\,.
 \label{homo}
\ee
The consequence of their inclusion on the CFT is
interpreted as fixing the initial and final states $|\psi_{\mathrm {i,f}}\rangle$ on which one computes the
expectation value on the rhs of \eqref{partition}. Our second observation is ii)   in Lorentzian signature the $z=\infty$
surface is a Killing horizon and therefore
an additional boundary where the bulk field needs to be specified for having
a well posed Dirichlet problem (see fig. \ref{contributions}(a)). These two observations
turn out to be related
to the fact that a second condition is required to fully
fix the bulk-boundary propagator {\sf K} (recall that in Euclidean space demanding regularity in the bulk
implies $K\to0$ when $z\to\infty$).
The remaining condition on {\sf K}  imposed at the horizon ($z=\infty$) is best expressed
in terms of Fourier modes as purely ingoing waves (exponentially
decaying) for timelike (spacelike) momenta, this is a well known problem for QFT in curved
spacetimes and amounts to the choice of vacuum. The incorporation of
normalizable (timelike) modes induce an outgoing component from the horizon
which is naturally interpreted as an excitation (see \cite{GKP,bala,bala2,balagid}
and \cite{vaman,kv,star,herson} for related work).

We are interested in computing two point correlation functions
on the dual field theory, to this end we need the on-shell action for a
scalar field to quadratic order
\be
S=-\frac12\int d\mathbf{x}dz
\sqrt{-g}\left(g^{\mu\nu}\partial_\mu\phi\,\partial_\nu\phi+m^2\phi^2\right)\,.
\label{freefi}
\ee
Integrating by parts and evaluating on shell, the contribution
from the conformal boundary is given by (see \cite{star} for a discussion on the
horizon contribution)
\be
S[\phi_0]=\frac12\int
d\mathbf{x}\left[\sqrt{-g}g^{zz}\phi(\mathbf{x},z)\,\partial_z\phi(\mathbf{x},z)\right]_{z=0}
\label{action scalar}\,.
\ee
Inserting \eqref{phi} into this expression gives the on-shell action as a functional of
the boundary data $\phi_0$
\be
S[\phi_0]=\frac12\int d\mathbf{y}d\mathbf{y'} \phi_0(\mathbf{y})\,\Delta(\mathbf{y},\mathbf{y'})\,\phi_0(\mathbf{y'})
\label{onshact}
\ee
where
\be
\Delta(\mathbf{y},\mathbf{y'})=\int
d\mathbf{x}\left[\sqrt{-g}g^{zz}{\sf K}(\mathbf{x},z\mid
\mathbf{y})\partial_z{\sf K}(\mathbf{x},z\mid
\mathbf{y'})\right]_{z=0}\,.\label{Delta}
\ee
Taking into account \eqref{K boundary condition} and \eqref{KfromG} in \eqref{Delta} one obtains
\bea
\label{Delta-fromG}
\Delta(\mathbf{y},\mathbf{y'})&\sim&\left[\sqrt{-g}g^{zz}\partial_z{\sf
K}(\mathbf{y},z\mid\mathbf{y'})\right] _{z=0}\\
&\sim&\lim_{z,z'\rightarrow0}(\sqrt{-g}g^{zz})(\sqrt{-g}g^{z'\!z'})\frac{\partial^2}{\partial z\,\partial{z'}}{\sf
G}(\mathbf{y},z\mid\mathbf{y'},z')\,.\label{DeltafromG}
\eea
This relation has been used to relate, in the semiclassical limit, the two point function
to the geodesics of the geometry \cite{shenker}.

Summarizing,
the two point function for an operator $\cal O$ dual to the bulk field $\phi$ is
obtained from the on shell action as
\be
\langle \psi_{\mathrm f}| \mathcal{O(\mathbf{y})}\mathcal{O(\mathbf{y'})}|\psi_{\mathrm i}\rangle=-i\frac{\delta^2
S[\phi_0]}{\delta\phi_0(\mathbf{y})\,\delta\phi_0(\mathbf{y'})}=-i\Delta^{\mathrm {i,f}}(\mathbf{y},\mathbf{y'})\,.
\label{2ptf}
\ee
The initial and final states $\psi_{\mathrm {i,f}}$ on the lhs encode to the ambiguity in adding
a normalizable solution to \eqref{phi}, in the following section we will show explicitly how they
manifest in \eqref{Delta}.

\subsection*{AdS global coordinates}

The recipe for obtaining QFT correlators from gravity computations
involves evaluating bulk quantities at the conformal boundary, as
might be suspected from \eqref{asympfi}, \eqref{K boundary condition}
and \eqref{Delta} the evaluation leads to singularities and
therefore requires a regularization. We will discuss in what follows
how this is done in the AdS global coordinate system since the
wormhole case we will discuss later coincides in its asymptotic
region with that coordinate system and will therefore be regularized
in the same way. Along the way we will show how the specification of
the initial and final states $\psi_{\mathrm {i,f}}$ appear in the
computation.

The regularization of \eqref{2ptf} is performed  imposing the boundary data at some finite distance
in the bulk and taking the limit to the boundary at the end of the computations (see \cite{rastelli} for
a subtlety when taking the limit). The AdS$_{d+1}$ manifold is fully covered by the so called global coordinates
where the metric takes the form
\be
ds^2=R^2\left[-\frac{dt^2}{1-x^2}+\frac{dx^2}{(1-x^2)^2}+\frac{x^2}{1-x^2}d\Omega^2_{d-1}\right]
\ee
where we have changed variables to $x=\tanh\rho$ from the standard radial $\rho$ variable
to map the conformal boundary to $x=1$.

We impose the boundary data at a finite distance $x_\epsilon=1-\epsilon$, therefore
consistency demands that
\be
 \lim_{x\to x_\epsilon}{\sf K}(t,{\Omega},x\mid t',{\Omega'},x_\epsilon)=\frac{\delta(t-t')
 \delta({\Omega}-{\Omega'})}{\sqrt{g_{_\Omega}}}\, .
 \label{Kbc}
\ee
and {\sf K} regular in the interior. The boundary-bulk propagator $\sf K$ satisfying \eqref{Kbc} can be obtained
from the Klein-Gordon equation solutions ${\sf \phi}(t,{\Omega},x)=e^{-i\omega
t}Y_{lm}({\Omega})f_{l\omega}(x) $ as
\be
{\sf K}(t,{\Omega},x\mid t',{ \Omega'},x_\epsilon)=\int_{-\infty}^{\infty}
\frac{d\omega}{2\pi}\sum_{lm}e^{-i\omega
(t-t')}Y_{lm}({\Omega})Y_{lm}^\ast({\Omega'})f_{l\omega}(x)
\label{propfeyn}
\ee
if we normalize $f_{l\omega}(x_\epsilon)=1$\footnote{The spherical
harmonics  $Y_{lm}$ on the $d-1$ sphere satisfy $\nabla^2 Y_{lm}=-q^2Y_{lm}$ with
$q^2=l(l+d-2),\,l=0,1,\ldots$}. For later comparison we quote the differential
equation satisfied by  $f_{l\omega}(x)$
\be
(1-x^2)\frac{d^2 f_{l\omega} }{dx^2}+\frac{d-1-x^2}{x}\frac{d
f_{l\omega}}{dx}+\left(\omega^2-\frac{q^2}{x^2}-
\frac{m^2R^2}{1-x^2}\right)f_{l\omega}=0\,.
\label{eqdiff2}
\ee
The solution to this equation is
a linear combination of two hypergeometric functions, but one of
them diverges as $x\rightarrow0$ so regularity in the bulk demands to discard it.
The properly normalized regular solution reads
\be
f_{l\omega}(x)=\frac{x^{-\frac{d}{2}+\nu +1}\left(1-x^2\right)^{\frac{1}{2}\Delta_+}}
{(1-\epsilon)^{-\frac{d}{2}+\nu+1} ((2-\epsilon ) \epsilon )^{\frac{1}{2}\Delta_+}}
   \frac{ \, _2F_1\left(\frac{1}{2} (\mu +\nu -\omega +1),\frac{1}{2} (\mu +\nu +\omega +1);\nu +1;x^2\right)}
   {\, _2F_1\left(\frac{1}{2} (\mu +\nu -\omega
   +1),\frac{1}{2} (\mu +\nu +\omega +1);\nu +1;(1-\epsilon)^2\right)}
\label{solglobal}
\ee
here $_2F_1$ is Gauss hypergeometric function with $\mu$ given by \eqref{lambda} and
$\nu=\sqrt{(\frac{d-2}{2})^2+q^2}$, the symmetry of
the hypergeometric function in its first two arguments  implies that $ f_{l\omega}(x)=f_{l\,-\omega}(x)$.
The asymptotic behavior of the solution \eqref{solglobal} near the boundary look
\be
f_{l\omega}(x)\sim C_+\, (1-x)^{\frac12{\Delta_+}}+C_-\,(1-x)^{\frac12\Delta_-}\,,
\label{asympglobal}
\ee
where $\Delta_\pm$ are given by \eqref{lambda} and $C_{\pm}=C_{\pm}(\mu,\nu,\omega)$.
In Lorentzian signature the KG operator posses normali\-zable
solutions, these appear for particular values of $\omega$
given by \cite{BF,bala,Avis}
\bea
 \omega_{nl}&=&\pm(2n+\nu+\mu+1)\nn\\
  &=&\pm(2n+l+\Delta_+),~~~~~~n,l=0,1,2\ldots\,,
\label{polos}
\eea
or stated otherwise, these are the frequencies for which $C_-=0$\footnote{See \cite{BF} for an
alternative quantization condition for $0\le\mu\le1$ and \cite{wittenDT} for its interpretation in
the AdS/CFT context.}. The discreteness of the spectrum manifests the ``box'' character
of AdS and from the dual perspective arises from the compactness of $S^3$. The quantization of the states \eqref{polos}
in the bulk is interpreted as dual to the QFT states defined on the $S^3\times \mathbb R$ conformal boundary
of AdS.
\begin{figure}
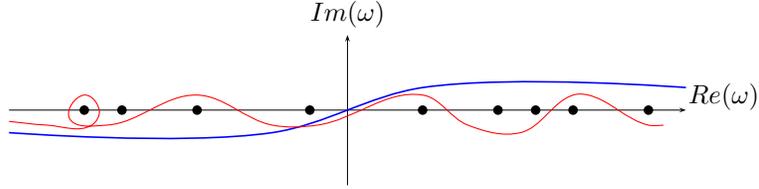

\vspace{2cm}
\hspace{8cm}
\psline[linewidth=.2pt]{->}(0,-1)(0,1)
\psline[linewidth=.2pt]{->}(-4.5,0)(4.5,0)
\psdots*(-3.5,0)(-3,0)(-2,0)(-0.5,0)(1,0)(2,0)(2.5,0)(3,0)(4,0)
\pscurve[linewidth=.6pt,linecolor=blue](-4.5,-0.3)(-1,-0.3)(1,0.3)(4.5,0.3)
\rput[bI](0,1.1){$Im(\omega)$}
\rput[bI](5,0){$Re (\omega)$}
\pscurve[linewidth=.4pt,linecolor=red](-4.5,-0.15)(-4,-0.2)(-3.4,-0.2)(-3.3,0)(-3.5,0.2)(-3.7,-0.1)(-3.2,-0.2)
(-2,0.2)(-1,-0.2)(-0.4,-0.2)(1,0.2)(1.6,-0.2)(2.3,-0.3)(3,0.2)(3.2,0.2)(4,-0.2)(4.2,-0.2)
\vspace{1cm}
\caption{{\bf Contours in $\bf \omega$ complex plane:} when performing the $\omega$ integration in \eqref{deltatt}
any arbitrary contour (depicted as red)  can be deformed to be the Feynman
contour (depicted in blue) plus contributions from encircling the poles \eqref{polos}. The encircling of positive (negative) frequency
poles fix the initial (final) states $\psi_{\mathrm {i,f}}$ in \eqref{2ptf}.}
\label{Contour}
\end{figure}

The two point correlation functions for the dual QFT operators are obtained by plugging
\be
\phi(t,\Omega,x)=\int dt'd\Omega'\sqrt{g_{_{\Omega'}}}\,{\sf K}(t,\Omega,x\mid
t',\Omega',x_\epsilon)\,\phi_0(t',\Omega') \label{identity}
\ee
into the action \eqref{freefi}, note that we have not included any normalizable
solution to \eqref{identity} (see next paragraph). The on-shell
action leads to a boundary term evaluated at $x_\epsilon$ (see \eqref{action scalar}-\eqref{Delta}) and the
regularized expression for
$\Delta(t,\Omega\mid t',\Omega')$ in \eqref{onshact} is therefore written as\footnote{Following \cite{rastelli},
when calculating \eqref{delta glob} we first compute the $x$-derivative and afterwards
we take the $\epsilon\to 0$ limit.}
\bea
  \Delta_{\sf reg}(t,\Omega|t',\Omega')&=&-\frac{1}{\sqrt{g_{_{\Omega}}}}\left[\sqrt{-g}g^{xx}\,\partial_x{\sf K}
  (t,\Omega,x\mid t',\Omega',x_\epsilon)\right] _{x=x_\epsilon}\nn\\
  &=&-\int\frac{d\omega}{2\pi}\,   e^{-i \omega (t- t')}
  \sum_{lm}Y_{lm}(\Omega)Y_{lm}^\ast(\Omega')
  \left[\frac{x^{d-1}}{(1-x^2)^{\frac{d-2}2}} \partial_x f_{l\omega}(x)\right]_{x=x_\epsilon}\nn\\
 &=& -\sum_{lm}Y_{lm}(\Omega)Y_{lm}^\ast(\Omega')\int \frac{d\omega}{2\pi}\,
 e^{-i \omega (t- t')}\left[\frac{x^{d-1}}{(1-x^2)^{\frac{d-2}2}} \partial_x f_{l\omega}(x)\right]_{x=x_\epsilon}\,,
 \label{delta glob}
\eea
where the first line comes from \eqref{Delta} taking into
account \eqref{Kbc}.

Some comments regarding \eqref{delta glob}: when taking the
$\epsilon\to 0$ limit, the expression in the last line turns out to be
ambiguos due to the existence of simple poles, located at
\eqref{polos}, along the  $\omega$-integration contour\footnote{Their origin
can be traced to having normalized $f_{l\omega}(x_\epsilon)=1$.}.
These poles manifest the existence of normalizable
solutions in the bulk (see \eqref{homo}) and therefore in order to define the
$\omega$-integration we need to give a prescription for
bypassing the poles. The choice of contour is traditionally
understood as the choice between advanced/retarded/Feynman Green
function, we will choose to work with the Feynman one in the following.
We now call the attention to the observation, pointed out in
\cite{Skenderis}, about the relation between contours in the complex
$\omega$-plane and choices of normalizable solutions. The
observation is simple, any particular choice of contour is
equivalent by deformation to choosing the Feynman contour plus
contributions from encircling the poles \eqref{polos}. Therefore
the ambiguity in the expression \eqref{identity} arising from the
addition of arbitrary normalizable modes translates into a choice of
contour in the complex $\omega$-plane (see fig. \ref{Contour}). The
Feynman contour choice naturally leads to time ordered correlators, and
the encircling of positive (negative) normalizable modes fix the initial
(final) state $\psi_{\mathrm {i,f}}$ in the lhs of \eqref{2ptf}. Choosing the
retarded contour as reference should be interpreted as giving rise to
response functions instead of correlation functions.
  Summarizing, the states are interpreted as created from a single fundamental
one $|\psi_0\rangle$ associated to the reference integration contour choosen.

The $\epsilon\to0$ limit of the expression inside the brackets in \eqref{delta glob}
also shows several poles in $\epsilon$ both analytic and non-analytic. The physical
result is obtained by renormalizing
the boundary data taking into account the asymptotic behavior in the radial direction
(see \eqref{asympglobal}), in the present case it amounts to rescale $\phi_0$ as
(see \cite{eddie,rastelli,mcg})
\be
\phi_0(t,\Omega)=\epsilon^{\frac12\Delta_-}\phi_{\sf ren}(t,\Omega)\,.
\label{renorm}
\ee
Moreover, since eventually we are interested in correlation functions for separated points, (contact)
terms proportional to positive integer powers of $q^2$ are dropped.
The finite term in the $\epsilon\to0$
limit reads
\bea
\Delta_{\sf ren}(t,\Omega| t',\Omega')&\equiv&\lim_{\epsilon \to0}\epsilon^{\Delta_-}\,\Delta_{\sf reg}(t,\Omega| t',\Omega')\nn\\&=&\sum_{lm}Y_{lm}(\Omega)Y_{lm}^\ast(\Omega')
\int \frac{d\omega}{2\pi}\,e^{-i \omega (t-t')}\nn\\
 &&\times\frac{\Delta_+}{2^{\Delta_-}}\frac{\Gamma (1-\mu )}{\Gamma (1+\mu)}
 \frac{\Gamma \left(\frac{1}{2}(-\omega+\nu +\mu +1)\right)\Gamma \left(\frac{1}{2} (\omega+\nu+\mu +1)\right)}
{\Gamma \left(\frac{1}{2} (-\omega+\nu -\mu +1)\right)
 \Gamma \left(\frac{1}{2} (\omega+\nu -\mu +1)\right) }\,.
 \label{deltatt}
\eea
The numerator of this last expression shows explicitly the appearance of poles along
the integration contour precisely at frequencies $\omega_{nl}$ given by \eqref{polos}. The specification
of a contour $\cal C$ in the complex $\omega$ plane fixes the initial and final states $\psi_{\mathrm {i,f}}$
when compared to the standard Feynman one (see fig. \ref{Contour}).

From the discussion following \eqref{delta glob} it should be clear that
correlation functions computed on the QFT vacuum state are obtained
by choosing the standard Feynman contour for the $\omega$
integration in \eqref{deltatt}. Performing the $\omega$ integration we
obtain\footnote{Generically \eqref{polos} are the only divergencies in \eqref{deltatt},
special care must be taken for integer values of $\mu$. We will not discuss
the details of this in the present work since experience with the AdS/CFT correspondence has shown
that correlation functions do not change qualitatively in the integer limit.}
\bea
\Delta_{\sf ren}^F(t,\Omega|t',\Omega')&=&2i\frac{\Delta_+\,\Gamma(1-\mu)}{2^{\Delta_-}\Gamma(1+\mu)}\sum_{lm}Y_{lm}(\Omega)Y_{lm}^\ast(\Omega')\nn\\
&&\times\left[\sum_{n=0}^\infty\frac{(-1)^{n}}{n!}\frac{\Gamma(n+l+\Delta_+)}{\Gamma(n+l+\frac d2)\Gamma(-(n+\mu))}
e^{-i|t-t'|(2n+l+\Delta_+)}\right]
\eea
The sum over the residues can be computed analytically giving
\bea
\langle 0|T{\cal O}(t,\Omega){\cal O}(t',\Omega')|0\rangle&=&-i\Delta_{\sf ren}^F(t,\Omega|t',\Omega')=
-\frac{2\Delta_+}{2^{\Delta_-}\Gamma(\mu)}
 \sum_{lm}Y_{lm}(\Omega)Y_{lm}^\ast(\Omega')\frac{\Gamma(l+\Delta_+)}{\Gamma(l+\frac d2)}\nn\\
 &&\times\,\, e^{-i|t-t'|(l+\Delta_+)}\,_2F_1\left(1+\mu,l+\Delta_+,l+\frac d2;e^{-2i|t-t'|}\right)\,. \label{2ptglobal}
\eea

\section{GKPW prescription for Lorentzian wormholes}
\label{gkpww}

Our goal in this section will be to extend the GKPW prescription to the case of multiple
timelike boundaries,
we will discuss the two boundaries case for simplicity. On general grounds
AdS/CFT suggests that the presence of two time-like boundaries
should be associated to the existence of two sets ${\cal O}^\pm$ of
dual operators corresponding to the two independent boundary
conditions $\phi^\pm_0$ that must be imposed on the field $\phi$
when solving the wave equation\footnote{Some authors have assumed
than the two dual field theories are decoupled because of the
disconnected structure of the boundary \cite{Azeyanagi}.}.

 We consider wormholes with (conformal) boundary topology of the
form ${\mathbb R}\times \Sigma $, with $\mathbb R$ representing time
and $\Sigma=\Sigma_++\Sigma_-$ the union of two (spatial) compact
disjoint copies $\Sigma_\pm$. The wormholes can be covered by a
single coordinate system $(x,t,\theta)$ where $x$ is the radial
holographic coordinate in the bulk and $(x_\pm,t,\theta)$ the
coordinates parametrizing the two boundaries $\mathbb
R\times\Sigma_\pm$.
\begin{figure}
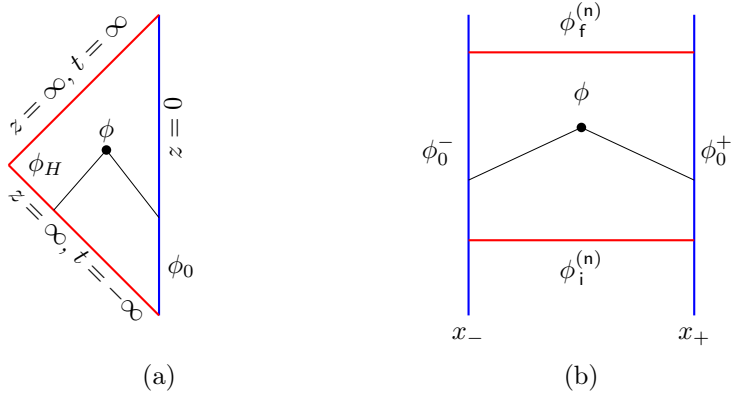

\vspace{2cm}
\hspace{5cm}
\psline[linewidth=.8pt,linecolor=blue](0,-2)(0,2)
\psline[linewidth=.8pt,linecolor=red](0,2)(-2,0)
\psline[linewidth=.8pt,linecolor=red](-2,0)(0,-2)
\psdots*(-0.7,0.2)
\psline[linewidth=.2pt](-0.7,0.2)(-1.4,-0.6)
\psline[linewidth=.2pt](-0.7,0.2)(0,-0.7)
\rput[bI](0,-3){(a)}
\rput[bI]{90}(0.3,0.5){$z=0$}
\rput[bI](0.3,-1.5){$\phi_0$}
\rput[bI](-1.5,-.15){$\phi_H$}
\rput[bI](-.7,.30){$\phi$}
\rput[bI]{45}(-1.1,1.1){$z=\infty$, $t=\infty$}
\rput[bI]{-45}(-1.2,-1.3){$z=\infty$, $t=-\infty$}
\hspace{4cm}
\psline[linewidth=.8pt,linecolor=blue](0,-2)(0,2)
\psline[linewidth=.8pt,linecolor=blue](3,-2)(3,2)
\psdots*(1.5,.5)
\psline[linewidth=.2pt](1.5,.5)(3,-0.2)
\psline[linewidth=.2pt](1.5,.5)(0,-0.2)
\psline[linewidth=.8pt,linecolor=red](0,-1)(3,-1)
\psline[linewidth=.8pt,linecolor=red](0,1.5)(3,1.5)
\rput[bI](1.5,-3){(b)}
\rput[bI](3,-2.4){$x_+$}
\rput[bI](0,-2.4){$x_-$}
\rput[bI](3.3,0){$\phi^+_0$}
\rput[bI](-.4,0){$\phi^-_0$}
\rput[bI](1.5,.8){$\phi$}
\rput[bI](1.5,-1.6){$\phi^{(\sf n)}_{\,\sf i}$}
\rput[bI](1.5,1.7){$\phi^{(\sf n)}_{\,\sf f}$}
\vspace{3cm}
\caption{(a) {\sf Poincare patch of Lorentzian AdS}: the value of the scalar field $\phi$ at any point in the bulk depends not only on the boundary value $\phi_0$
but also on  the value of the field at the future and past horizons $\phi_H$. ~~~~~~(b) {\sf Lorentzian wormhole geometry}:
the value of the scalar field $\phi$ at any point in the bulk depends not only on two boundary values $\phi^\pm_0$ (blue)
but also on normalizable modes in the bulk $\phi^{(\sf n)}$ (red). The dependence, in both pictures, on the normalizable modes (red lines)
correlates to a choice of the initial and final states $|\psi_{\mathrm {i,f}}\rangle$
and manifest as a choice of contour in the complex $\omega$ plane when computing the correlator  \eqref{deltatt}.}
\label{contributions}
\end{figure}
In the presence of two disconnected conformal boundaries we propose to write the bulk field in terms of the
boundary data $\phi_0^{\pm}(\mathbf{y})$ on each of the boundaries  as (see figure \ref{contributions})
\bea
\phi(\mathbf{y},x)&=&\int d\mathbf{y}' \,
{\sf K}^{i}(\mathbf{y},x|\mathbf{y}'  )\,\phi^{i}_0(\mathbf{y}' )\nn\\
&=&\int d\mathbf{y}' \,
\left[{\sf K}^{+}(\mathbf{y},x|\mathbf{y}'  )\,\phi^{+}_0(\mathbf{y}' )+ {\sf
K}^{-}(\mathbf{y},x|\mathbf{y}')\,\phi^{-}_0(\mathbf{y}')\right]
\,.
\label{phi2bdy}
\eea
here $\mathbf{y}=(t,\theta)$, and note that the
solution for given boundary data is not unique since in Lorentzian
signature normalizable solutions can be added to \eqref{phi2bdy}.
This ambiguity is resolved, as discussed in section \ref{GKPW}, when
a choice of contour in the frequency space $\omega$ of the kernel
$\sf K$ is given. Our method differs from the proposal developed in
\cite{balasu}: in that work only the $K^+$ bulk-boundary propagator
was discussed and its form was determined by demanding the absence
of cuts when extending the radial coordinate to complex values. The
prescription lead to the conclusion that $\phi_0^\pm$ were not
independent.

Consistency demands the bulk-boundary propagators ${\sf K}^{\pm}(\mathbf{y},x|\mathbf{y}')$
to solve the Klein-Gordon equation \eqref{box} with the following boundary conditions
 \bea
 \left.{\sf K}^+(\mathbf{y},x|\mathbf{y}')\right|_{x=x_+}&=&\delta(\mathbf{y}-\mathbf{y}'),~~~~~~~~
 \left.{\sf K}^+(\mathbf{y},x|\mathbf{y}')\right|_{x=x_-}=0\nonumber\\
 \left.{\sf K}^-(\mathbf{y},x|\mathbf{y}')\right|_{x=x_-}&=&\delta(\mathbf{y}-\mathbf{y}'),~~~~~~~~
 \left.{\sf K}^-(\mathbf{y},x|\mathbf{y}')\right|_{x=x_+}=0
 \label{condition2boundaries}\,,
\eea
These expressions completely determine the bulk-boundary propagators $\sf K^\pm$.
The on-shell action \eqref{freefi} results in two terms arising from the boundaries which take the form
\be
S=-\frac12\int d\mathbf{y}\left(\left[\sqrt{-g}g^{xx}\phi(\mathbf{y},x)\partial_x\phi(\mathbf{y},x)\right]_{x=x_+}
-\left[\sqrt{-g}g^{xx}\phi(\mathbf{y},x)\partial_x\phi(\mathbf{y},x)\right]_{x=x_-}\right)
\label{general two bdy action}
\ee
Inserting the solution
\eqref{phi2bdy} into \eqref{general two bdy action} one obtains
\bea
S[\phi_0]=-\frac12\int d\mathbf{y}\,\, d\mathbf{y}'\phi_0^i(\mathbf{y})\Delta_{ij}(\mathbf{y},\mathbf{y}')\phi_0^j(\mathbf{y}')
\label{two bdy on shell action}
\eea
with $i,j=+,-$ denoting the two boundaries and $\Delta_{ij}$ the generalization of \eqref{Delta}. Their
explicit forms are
\be
\Delta_{+i}(\mathbf{y},\mathbf{y}')=  \left[\sqrt{-g}g^{xx} \partial_x{\sf K}^i(\mathbf{y},x|\mathbf{y}')\right]_{x=x_+}\,,~~~~~
\Delta_{-i}(\mathbf{y},\mathbf{y}')=  -\left[\sqrt{-g}g^{xx} \partial_x{\sf K}^i(\mathbf{y},x|\mathbf{y}')\right]_{x=x_-}
\label{two bdy deltas}
\ee
As in Section \ref{GKPW} the two point functions of operators on the same boundary result
\be
\langle\psi_{\mathrm f}| {\cal O}^\pm(\mathbf{y}){\cal O}^\pm(\mathbf{y'})|\psi_{\mathrm i}\rangle\sim
-i\Delta_{\pm\pm}(\mathbf{y},\mathbf{y'})\label{samebdy}
\ee
and the correlators between operators on opposite boundaries read
\be
\langle\psi_{\mathrm f}| {\cal O}^\pm(\mathbf{y}){\cal O}^\mp(\mathbf{y'})|\psi_{\mathrm i}\rangle\sim-i\Delta_{\pm\mp}(\mathbf{y},\mathbf{y'})\,.
\label{opposbdy}
\ee
The generalization of the expressions \eqref{KfromG} and \eqref{DeltafromG} to backgrounds
with two  boundaries take the form
\be\label{K-fromG-2bound}
{\sf K}^i(\mathbf{y},x\mid\mathbf{y}')=
\lim_{x'\rightarrow x^i}\sqrt{-g}\, g^{x'x'}\partial_{x'}{\sf G}(\mathbf{y},x\mid\mathbf{y}',x'),
\ee
which gives
\be\label{Delta-fromG-2bound}
\Delta_{ij}(\mathbf{y},\mathbf{y}')\sim\lim_{x\rightarrow x^i,~ x'\rightarrow x^j
}(\sqrt{-g}g^{xx})
(\sqrt{-g}g^{x'\!x'})\frac{\partial^2}{\partial_x\partial_{x'}}{\sf G}(\mathbf{y},x\mid\mathbf{y}',x')\,.
\ee
Note that the $\Delta_{\pm\mp}$ correlation function involves
in the semiclassical limit a geodesic through the bulk connecting two points, one at each boundary.

\subsection*{$AdS_2$ Lorentzian strip}

We will apply in this section the prescription developed above to Lorentzian AdS$_2$
reobtaining previous results. The AdS$_2$ Lorentzian metric can be written as
\be
 ds^2=R^2\left[-\frac{dt^2}{1-x^2}+\frac{dx^2}{(1-x^2)^2}\right]\,,
\label{ads2}
\ee
the timelike boundaries are located at $x=\pm1$ and the $(t,x)$ coordinate system covers the whole spacetime.
To find the bulk-boundary
propagators $\mathsf K^{\pm}$ in \eqref{phi2bdy} we propose
\be
{\sf K^{\pm}}(t,x)=\int_{-\infty}^{\infty}\frac{d\omega}{2\pi}e^{-i\omega t}f_\omega^{\pm}(x)\,.
\ee
Inserting into the KG equation \eqref{box} we obtain  the following differential equation for $f_\omega$
\be
 (1-x^2)\frac{d^2 f_\omega^{\pm}(x)}{dx^2}-{x}\frac{d
 f_\omega^{\pm}(x)}{dx}+\left(\omega^2-
 \frac{m^2R^2}{1-x^2}\right)f_\omega^{\pm}(x)=0\,.
 \label{efads2}
\ee
The solution to \eqref{efads2} can be written in terms of generalized Legendre Polynomials as
\be
 f_{\omega}^{\pm}(x)=(1-x^2)^{\frac{1}{4}}[a_{\omega}^{\pm}\,P^\mu_\nu(x)+b_{\omega}^{\pm}\,Q^\mu_\nu(x)]
 \label{fm}
\ee
with  $\mu=\sqrt{\frac1{4}+m^2R^2}$,  $\nu=\omega
-\frac12$ and $a_\omega^{\pm}, b_\omega^{\pm}$ arbitrary constants which get fixed when we impose the conditions
\eqref{condition2boundaries}. The conditions translate into\footnote{As discussed in section \ref{GKPW}
the boundary data is imposed at a finite distance $x=\pm x_\epsilon$ where $x_\epsilon=1-\epsilon$.}
\be
f_{\omega}^\pm(\pm x_\epsilon)=1\,,~~~~~~~~
f_{\omega}^\pm(\mp x_\epsilon)=0\,.
\label{norm}
\ee
The solutions to \eqref{norm} read
\bea
f_{\omega}^{+}(x)=\frac{(1-x^2)^{1/4}}{(1-x_\epsilon^2)^{1/4}}\,\frac{Q^\mu_\nu(x)P^\mu_\nu(-x_\epsilon)-Q^\mu_\nu(-x_\epsilon)P^\mu_\nu(x)}
{Q^\mu_\nu(x_\epsilon)P^\mu_\nu(-x_\epsilon)-Q^\mu_\nu(-x_\epsilon)P^\mu_\nu(x_\epsilon)}\\
f_{\omega}^{-}(x)=\frac{(1-x^2)^{1/4}}{(1-x_\epsilon^2)^{1/4}}\,\frac{Q^\mu_\nu(x)P^\mu_\nu(x_\epsilon)-Q^\mu_\nu(x_\epsilon)P^\mu_\nu(x)}
{Q^\mu_\nu(-x_\epsilon)P^\mu_\nu(x_\epsilon)-Q^\mu_\nu(x_\epsilon)P^\mu_\nu(-x_\epsilon)}\,.
\eea
Analyzing the asymptotic behavior near the boundary  in \eqref{fm} one finds
normalizable modes for
\be
\omega_n=\pm\left(n+\mu+\frac12\right),~~~~~~n=0,1,2\ldots ~~~
 \mathrm {and}~~~~\frac{b_{\omega Q}}{a_{\omega Q}}=-\frac{2\tan\pi\mu}{\pi}  \label{polosads2}
\ee
The renormalized  $\Delta_{ij}$  functions disregarding contact terms result
\bea
\Delta_{\mathsf {ren}_{\pm\pm}}(t,t')&=&\mp\frac{2^{\Delta_-}}{2\pi}\frac{\Gamma (1-\mu )}{\Gamma (1+\mu )}\int
\frac{d\omega}{2\pi} e^{-i\omega(t-t')}
 \Gamma \left(\frac12+\mu-\omega \right) \Gamma \left(\frac12+\mu+\omega\right) \cos (\pi  \omega )
 \label{act}\\
\Delta_{\mathsf
{ren}_{\pm\mp}}(t,t')&=&\mp\,\frac{2^{\Delta_-}}{\Gamma (\mu
)^2}\int \frac{d\omega}{2\pi}e^{-i\omega(t-t')} \Gamma
\left(\frac12+\mu-\omega\right) \Gamma \left(\frac12+\mu+\omega\right) \,, \label{actionterms}
\eea
as before, the integrands in these   expressions show
poles in $\omega$ at the values given by \eqref{polosads2}.

The  integrals \eqref{act}-\eqref{actionterms} can be computed  using the
residue theorem once a contour in the complex plane is chosen. For
the Feynman contour we obtain
\bea
\Delta_{\mathsf
{ren}_{\pm\pm}}^F(t,t')&=&\mp\frac{i^{\Delta_-
-\Delta_+}}{8^{\Delta_+}\pi^{\frac12}}
\frac{\Gamma(\frac12+\mu)}{\Gamma(\mu)
\sin^{2\Delta_+} \left(\frac{t-t'}{2}\right)}\nn\\
\Delta_{\mathsf {ren}_{\pm\mp}}^F(t,t')&=&\mp\frac{8^{\Delta_-}i}{4}
\frac{\Gamma(1+2\mu)}
{\Gamma(\mu)^2\cos^{2\Delta_+} \left(\frac{t-t'}{2}\right)}
\eea
The vacuum expectation values between operators on the same and opposite boundaries result (cf. \cite{Azeyanagi})
\bea
\langle 0|T
{\mathcal O}^\pm(t){\mathcal O}^\pm(t')|0\rangle&=&\pm\frac{4^{\Delta_- }i^{2\Delta_-}}{8^{\Delta_+}}\frac{\Gamma(2\mu)}
{\Gamma(\mu)^2\sin^{2\Delta_+}(\frac{t-t'}{2})},\\
\langle 0| T{\mathcal O}^\pm(t){\mathcal
O}^\mp(t')|0\rangle&=&\mp\frac{8^{\Delta_-}}{4}\frac{\Gamma(1+2\mu)}
{\Gamma(\mu)^2\cos^{2\Delta_+}(\frac{t-t'}{2})}. \label{ads2++}
\eea The first line gives the result for operators on the same
boundary and has the expected conformal behavior $|
t-t'|^{-2\Delta_+}$ when the operators approach each other. The
second line corresponding to operators located on different
boundaries becomes singular for $t=t'+(2n+1)\pi$, $n\in\mathbb Z
$, this singularity reflects the existence of causal (null) curves
connecting the boundaries and it has been argued that their
existence hints for an interaction between the two sets of degrees
of freedom ${\cal O}^\pm$ \cite{balasu}\footnote{\label{spatial}
This  interpretation assumes that the boundary spatial foliations
$\Sigma_+^{(t)}$ and $\Sigma_-^{(t')}$  should be identified as
being the same Cauchy surface $\tilde\Sigma$ of a single base
manifold where the dual field theory lives in.}.
 The
observed  periodicity in time relates to a peculiar property of AdS,
this  is the convergence of null geodesics when passing to the
universal cover and can be also understood as a consequence of the
eigenmodes \eqref{polosads2} being equally spaced (see
\cite{Avis}-\cite{hawkingellis}).

In the massless case ($\mu=\frac12$) the two point functions take the form
\bea
\langle 0|T
{\mathcal O}^\pm(t){\mathcal O}^\pm(t')|0\rangle=\pm\frac{1}{8\pi\sin^2\left(\frac{t-t'}{2}\right)}\,,~~~
~\langle 0|T {\mathcal O}^\pm(t){\mathcal O}^\mp(t')|0\rangle=\mp\frac{1}{4\pi\cos^2\left(
\frac{t-t'}{2}\right)}
\eea

\subsection*{Wormhole}

We now turn to the analysis of two point functions in a wormhole background, this is, a spacetime geometry with two conformal boundaries connected
through the bulk. We will work with a toy model wormhole, which permits analytical treatment, consisting on a static geometry that
connects two asymptotically AdS regions with base manifolds of the form
$H^3$ or $S^1\times H^2$, with
$H^n$  a  $n$-dimensional (quotiented) hyperbolic space. The geometry
does not contain horizons anywhere, and the two asymptotic regions are causally
connected. The spacetime was found  as a solution of Einstein-Gauss-Bonnet gravity, which
in $d+1=5$ dimensions takes the form
\be
S_5=\kappa\int\epsilon_{abcde}\left(R^{ab}R^{cd}+\frac{2}{3l^2}R^{ab}e^ce^d+\frac{1}{5l^4}e^ae^be^ce^d\right)e^e\nn\label{EGB}\,,
\ee
here $R^{ab}=d\omega^{ab}+\omega^a_f\omega^{fb}$ is the curvature two form for the spin connection $\omega^{ab}$, and $e^a$ is the vielbein.
The $(d+1)$-dimensional wormhole metric found in \cite{Dotti} reads
\bea
  ds^2&=&R^2\left[-\cosh^2\!\rho\,{dt^2}+{d\rho^2}+\cosh^2\!\rho\,{d\tilde\Sigma^2_{d-1}}\right]\nn\\
 &=&R^2\left[-\frac{dt^2}{1-x^2}+\frac{dx^2}{(1-x^2)^2}+\frac{d\tilde\Sigma^2_{d-1}}{1-x^2}\right]
 \label{wh}
\eea
where $d\tilde\Sigma^2_{d-1}$ is a  constant negative curvature metric on the compact
base manifold $\Sigma^2_{d-1}$,  two disconnected conformal boundaries are located at $x=\pm1$.

To construct the boundary to bulk propagators ${\sf K}^\pm$ discussed above we
propose
\be
{\sf K}^\pm(t,x,\theta|t' ,\theta')=
\int_{-\infty}^{\infty}
\frac{d\omega}{2\pi}\sum_{Q}e^{-i\omega (t-t')}\,Y_{Q}(\theta)Y_{Q}^\ast(\theta')f_{ \omega Q}^\pm(x)
\label{ansatz}
\ee
where $Y_Q(\theta)$ are the spherical
harmonics on the base manifold\footnote{The spherical harmonics satisfy $\nabla_\Sigma ^2Y_Q=-Q^2\,Y_Q$ and a compact
manifold without boundaries has $Q^2\ge0$.}.
Inserting (\ref{ansatz}) into the KG equation (\ref{box}) one finds that $f_{ \omega Q}$ satisfies
\be
 (1-x^2)\,\frac{d^2 f_{\omega Q }^{\pm}(x)}{dx^2} +(d-2)\,x \frac{d f_{\omega Q }^{\pm}(x)}{dx}+
\left[ (\omega^2-Q^2)-\frac{m^2R^2}{1-x^2}\right]f_{\omega Q
}^{\pm}(x)=0 \label{eqdiffworm}\,.
\ee
The solutions to
\eqref{eqdiffworm} can be written in terms of generalized Legendre
polinomials as \cite{diego}
\be f_{\omega Q
}^{\pm}(x)=(1-x^2)^{\frac{d}{4}}[a_{\omega Q}^{\pm}\,P^\mu_\nu(x)+b_{\omega
Q}^{\pm}\,Q^\mu_\nu(x)] \label{solution}
\ee
where $\mu$ is given by  \eqref{lambda} and
\bea
\nu=\varpi-\frac12=\sqrt{\left(\frac{d-1}{2}\right)^2+\omega^2-Q^2}-\frac12\,.
\label{nu}
\eea
$a_{\omega Q}^{\pm},b_{\omega Q}^{\pm}$ in \eqref{solution} are
constant coefficients which get fixed when we impose the conditions
\eqref{condition2boundaries}, these are
\be
f_{\omega Q}^\pm(\pm x_\epsilon)=1\,,~~~~~~~~
f_{\omega Q}^\pm(\mp x_\epsilon)=0\,.
\label{normalization}
\ee
The solutions for \eqref{solution} satisfying \eqref{normalization} are
\bea
f^+_{\omega Q}(x)&=&\frac{\left(1-x^2\right)^{\frac{d}{4}}} {\left(1-x_\epsilon ^2\right)^{\frac{d}{4}}}\,
\frac{ P_{\nu }^{\mu }(x)Q_{\nu }^{\mu }(-x_\epsilon)-P_{\nu }^{\mu }(-x_\epsilon ) \,Q_{\nu }^{\mu
   }(x)}{P_{\nu }^{\mu }(x_\epsilon )Q_{\nu }^{\mu
   }(-x_\epsilon)-P_{\nu }^{\mu }(-x_\epsilon) Q_{\nu }^{\mu }(x_\epsilon )}\nn\\
f^{-}_{\omega Q}(x)&=&\frac{\left(1-x^2\right)^{\frac{d}{4}}}{ \left(1-x_\epsilon^2\right)^{\frac{d}{4}}}\,
\frac{ P_{\nu }^{\mu }(x)Q_{\nu }^{\mu }(x_\epsilon )-P_{\nu }^{\mu }(x_\epsilon ) Q_{\nu }^{\mu
   }(x)}{{P_{\nu }^{\mu }(-x_\epsilon)Q_{\nu }^{\mu
   }(x_\epsilon )-P_{\nu }^{\mu }(x_\epsilon ) Q_{\nu }^{\mu }(-x_\epsilon)}}\label{solworm}\,,
\eea
The possibility for two independent bulk-boundary propagators arises from the fact that the base manifold $\tilde\Sigma$
never shrinks to zero size inside the bulk (see \eqref{wh}) and therefore regularity in the interior imposes no constraint
in the solutions \eqref{solution}.
Normalizable modes  appear for
\be
 \omega_{nQ}=\pm
\sqrt{\left(\mu+\frac12+n\right)^2+Q^2-\left(\frac{d-1}{2}\right)^2}\,,~~~~~n=0,1,\ldots,~~~
 \mathrm {and}~~~~\frac{b_{\omega Q}}{a_{\omega Q}}=-\frac{2\tan\pi\mu}{\pi} \label{poloswh}
\ee
For these frequencies the  index $\nu$ in  \eqref{nu} takes the value $\nu_n=\mu+n$ and the resulting solution
becomes normalizable.
The two point functions \eqref{samebdy}-\eqref{opposbdy} between operators ${\cal O}_{\pm}$ on the same
boundary and opposite boundaries take the form
\bea
\langle \psi_{\mathrm {f}}|{  T}{\mathcal O}^\pm(t,\theta){\mathcal O}^\pm(t',\theta')|\psi_{\mathrm {i}}\rangle
&=&\pm\,i \frac{2^{\Delta_-}d}{\pi2^d}\frac{\Gamma(1-\mu)}{\Gamma(1+\mu)}\,\sum_{ Q}Y_Q(\theta)Y_{Q}^\ast(\theta')\nn\\
&&\quad\quad\times\int\frac{d\omega}{2\pi}e^{-i\omega(t-t')}\Gamma(\frac12+\mu-\varpi)\Gamma(\frac12+\mu+\varpi) \cos(\pi\varpi)
\label{w++}\\
\langle \psi_{\mathrm {f}}|T{\mathcal O}^\pm(t,\theta){\mathcal O}^\mp(t',\theta')|\psi_{\mathrm {i}}\rangle
&=&\pm\,i \frac{2^{\Delta_-}}{2^{d-1}}\frac{1}{\Gamma (\mu )^2}\,\sum_{Q}Y_Q(\theta)Y_{Q}^\ast(\theta')\nn\\
&&\quad\quad\times\int\frac{d\omega}{2\pi}e^{-i\omega(t-t')} \Gamma(\frac12+\mu-\varpi)\Gamma(\frac12+\mu+\varpi).
\label{w+-}
\eea
A few comments on these expressions: (i) in the large frequency limit $\varpi\sim \omega$ and the integrands in \eqref{w++}-\eqref{w+-} coincide
with those of AdS$_2$ (cf. \eqref{act}-\eqref{actionterms}), (ii) when a Feynman contour is chosen, a time ordered
product should be understood on the rhs of \eqref{w++}-\eqref{w+-} and (iii) although the Gamma functions $\Gamma(\frac12+\mu\pm\varpi)$ present
two branch cuts at $\omega=\pm\sqrt{Q^2-\left(\frac{d-1}{2}\right)^2}$ (see \eqref{nu}), the product in the integrands \eqref{w++}-\eqref{w+-} is free of them.

The correlation between operators inserted on opposite boundaries is non vanishing, and this result
has been explained in different ways depending on the context:
(i) as the result of being computing the correlator \eqref{w+-} on an entangled state of two  non-interacting boundary theories (BH context \cite{Eternal}) or (ii) as due
to an interaction between the theories defined on each of the boundaries (D1/D5 orbifold in \cite{balasu}). The crucial point
in both arguments was the absence/existence of causal connection between the asymptotic regions.

\section{Entanglement vs. Coupling}\label{holentanglement}

In this section we will review several thoughts regarding the
interpretation of the results \eqref{ads2++} and \eqref{w+-}. We would like to address the
issue of whether the results \eqref{ads2++} and \eqref{w+-} are the consequence of: (i) an interaction
between the two dual QFT theories or (ii) due to the correlators being evaluated on an entangled state or
(iii) both.

\subsection*{Entanglement Entropy}

The entanglement entropy $\mathsf S_{_{\mathrm A}}$ is a non local quantity (as opposed to correlation functions)
that measures how two subsystems A and B are correlated.
For a $d$-dimensional QFT, it is defined as the von Neumann entropy
of the reduced density matrix $\rho_A$ obtained when we trace out the degrees of freedom inside a
$(d-1)$-dimensional spacelike submanifold B which is the complement of A (see \cite{rt} for a review).

In \cite{Ryu, takaya} a proposal was made for a holographic formula for the entanglement entropy
of a CFT$_d$ dual to an AdS$_{d+1}$ geometry, it reads
\be
\mathsf S_{_{\mathrm A}}=\frac{\mathrm{Area}(\gamma_{_{\mathrm A}})}{4G_N^{(d+1)}}\,,
\label{entang}
\ee
where $\gamma_{_{\mathrm A}}$ is a $(d-1)$-dimensional minimal surface in AdS$_{d+1}$ whose boundary $\cal S$,
located at AdS infinity, coincides with that of A, this is ${\cal S}=\partial\gamma_{_{\mathrm A}}=\partial {\mathrm A}$ and
$G_N^{(d+1)}$ is the Newton constant in AdS$_{d+1}$. This formula, which assumes the supergravity
approximation of the full string theory, has been applied in the
AdS$_3$/CFT$_2$ setup showing agreement with known 2D CFT results \cite{rt}.

One can imagine applying a generalization of \eqref{entang} to the wormhole
geometry \eqref{wh} as follows\footnote{
See \cite{Hubeny} for an generalization of \eqref{entang} to the euclidean wormhole constructed in \cite{maldamaoz}.}:
the wormhole presents two disconnected spatial regions $\Sigma_\pm$ at which two
identical degrees of freedom are supposed to be living on. Imagine constructing
a codimension 1 closed surface $\cal S$ bounding a {\it small} region B on $\Sigma_-$,
experience from Wilson loops and brane embeddings show that as  B  remains small, the minimal surface will
be located near $x=-1$. As we gradually increase the size of
B, the minimal surface anchored on $\Sigma_-$ dips deeper into
the bulk and in the limit when B occupies all space B$\rightarrow \Sigma_-$
the boundary $\cal S$ collapses and the minimal surface gets
localized at the throat $x=0$ of the geometry, giving a non zero result
\be
\mathsf S_{_{\tilde\Sigma}}=\frac{\mathrm{Area}(\tilde\Sigma)}{4G_N^{(d+1)}}\,.
\label{entro}
\ee
This result should be understood as indicating that the quantum state in the dual QFT
described by the wormhole geometry is not separable, or stated otherwise, the result \eqref{w+-} can
be attributed to the wormhole geometry realizing an entangled state on the Hilbert space product
$\cal H=\cal H_+\otimes\,\cal H_-$ with the entropy \eqref{entro} resulting from integrating out $\cal H_-$.

\subsection*{Interaction between the dual copies}
\label{double boundary geos}

We will now argue that apart from the entanglement discussed above, the causal contact between the wormhole asymptotic boundaries
leads to an interaction term between the two sets of degrees of freedom living at each boundary. In particular we argue that
a coupling between the field theories will exist whenever the asymptotic regions are in causal contact.

We start quoting the partition function on the gravity side for the wormhole geometry
in the semiclassical limit, its form is
\be
\mathcal{Z}_{gravity}\left[\,\phi(\phi_0^+,\phi_0^-,{\cal C})\right]\sim\, e^{-\frac{i}{2}
\int d{\mathbf{y}}\,\, d\mathbf {y}'\phi_0^i(\mathbf{y})\Delta_{ij}(\mathbf{y},\mathbf{y}')\phi_0^j(\mathbf{y}')}
\label{Z-2Bs}
\ee
here ${\bf y}=(t,\theta)$  denote the boundary points and  $i,j=+,-$ refer to the asymptotic boundaries
where prescribed  data $\phi_0^\pm$ is given, the contour $\cal C$ fixes the normalizable solution in the bulk
and
the expression for $\Delta_{ij}$ is given by \eqref{two bdy deltas}.
According to the GKPW prescription the partition function \eqref{Z-2Bs} is the generating functional for ${\cal O}^\pm$
correlators, this is
\be\
\label{witten-gen-pm}
\mathcal{Z}_{gravity}\left[\phi(\phi_0^+,\phi_0^-,\cal C)\right]=\left\langle\,\psi_{\mathrm f}
\left|\,T
e^{i\int
d{\bf y}\,\phi_0^+({\bf y})
{\cal O^+({\bf y})} + i\int
d {\bf y}\,\phi_0^-({\bf y}){\cal O}^-({\bf y})}
\right|\,\psi_{\mathrm i}\right\rangle_{_{QFT}}\,,
\ee
where the observables ${\cal O}^\pm$  should be constructed as local functionals of the fundamental
fields $\Psi_\pm$ living on each boundary. These  fields  are assumed to describe independent degrees of
freedom: $[{\cal O}^+ , {\cal O}^-]=0$ on the same spatial slice (see footnote \ref{spatial}).

Consider the simplest situation corresponding to choosing $\cal C$ to be the Feynman contour,
this is, we are computing the vacuum to vacuum transition amplitude
on the field theory side. The rhs in \eqref{witten-gen-pm} can be written as
\be
\left\langle \psi_0 \left|\,T e^{i\int
d {\bf y}\,\phi_0^+({\bf y}){\cal O^+({\bf y})}\,+\,i \int
d {\bf y}\,
\phi_0^-({\bf y}){\cal O}^-({\bf y})}\right|\psi_0\right\rangle_{_{QFT}}=\mathsf {Tr}\left[\,\rho_{_{\psi_0}}\,T
e^{i\int
d {\bf y}\,\phi_0^+({\bf y}){\cal O^+({\bf y})} \,+\, i\int
d {\bf y}\,\phi_0^-({\bf y}){\cal O}^-({\bf y})}\right]
\label{whcorrel},
\ee
where the trace operation is performed over   a complete set of states  of the dual field theory Hilbert space
and $\rho_{_{\psi_0}}=|\psi_0\rangle\langle \psi_0 |$ is the density matrix associated   to the vacuum wormhole state.
This vacuum state belongs to the Hilbert space ${\cal H} = {\cal H}_+ \otimes {\cal H}_-$
and according to the  arguments reviewed in the last subsection, it is not separable as a single tensor product
$|\psi_0\rangle\ne|\psi_+\rangle\otimes|\psi_-\rangle$.

To analyze the possibility of interaction between the fields living at each boundary we
consider the system at finite temperature $T=\beta^{-1}$.
The absence of singularities in the Euclidean continuation implies
that the wormhole spacetime can be in equilibrium with a thermal
reservoir of arbitrary temperature, or stated otherwise, the
periodicity in Euclidean time is arbitrary. At thermal
equilibrium the field theory state is described by the Boltzmann distribution
$\rho_{_\beta}= e^{-\beta H}$, with $H=H_+[\Psi_+] + H_-[\Psi_-] +
H_{int}[\Psi_+, \Psi_-]$ the dual field theory Hamiltonian, and $H_{int}$
a possible coupling between the two identical sets of degrees of
freedom $\Psi_\pm$.

Let us see now that an interaction term $H_{int}$ should be present  in the Hamiltonian in order to avoid a contradiction.
The argument goes as follows: finite temperature correlation functions on the field theory side are obtained from the Euclidean rotation
of \eqref{whcorrel}, which reads
\be
\left\langle  e^{- \int
d {\bf y}\,\phi_0^+({\bf y}){\cal O^+({\bf y})}\, -\, \int
d {\bf y}\,
\phi_0^-({\bf y}){\cal O}^-({\bf y}) }\right\rangle_{{\beta
}} =
\mathsf{Tr}\left[\,\rho_{_\beta}\, e^{-\int d {\bf y}\, \phi_0^+({\bf y}) {\cal O}^+({\bf y})\,
 -\, \int d {\bf y}\,\phi_0^-({\bf y}){\cal O}^-({\bf y}) }\,\right]\,,\label{euclqft}
\ee
the field theory is defined on an Euclidean manifold with $S^1_\beta\times \tilde\Sigma$ topology (see footnote \ref{spatial}).
The AdS/CFT proposal equates this expression to the Euclidean continuation of the lhs in \eqref{witten-gen-pm},
the result is
\bea
\label{thermalAdSCFT}
\mathsf{Tr} \,\left[\, e^{-\beta H}\, \,e^{-\int d {\bf y}\,\phi_0^+({\bf y}) {\cal O^+({\bf y})} -
\int d {\bf y}\,\phi_0^-({\bf y}){\cal O}^-({\bf y})} \,\right] &=&
\mathcal{Z}_{Egravity}\left[\phi(\phi_0^+,\,\phi_0^-)\right]\sim e^{ -S_E[\phi_0^+ ,\,\phi_0^-]}\,,
\label{eucZ}
\eea
on the right $S_E[\phi_0^+ ,\,\phi_0^-]$ corresponds to the on shell scalar field action
on the Euclidean wormhole background, with time  compactified on a circle of radius $\beta$. Upon Euclidean
rotation and time compactification, the resulting geometry is cylinder like with topology  $S_\beta^1 \times I  \times \tilde\Sigma$,
the two boundaries at which one imposes the boundary data $\phi_0^\pm$  are located at the endpoints $x_\pm$ of the finite  interval
$I$\footnote{ Note the similarity with the geometry  studied in \cite{maldamaoz}.}. Note that in the Euclidean context the solution in the
interior (bulk) is completely specified by the boundary data, no normalizable solutions exist and  rotating the Feynman contour
to the imaginary time axis is straightforward and leads to a non-singular solution. The explicit expression for the rhs of \eqref{eucZ} is
\be\label{eucl-AdSCFT}
\mathcal{Z}_{Egravity}\left[\phi(\phi_0^+,\,\phi_0^-)\right]\sim\, e^{
-\frac12\int d\mathbf{y}\,\, d\mathbf{y}'\phi_0^i(\mathbf{y})\,\tilde\Delta_{ij}(\mathbf{y},\mathbf{y}')\,\phi_0^j(\mathbf{y}')},
\ee
where $\tilde{\Delta}_{ij}$ denote the Euclidean rotated bulk-boundary propagators \eqref{two bdy deltas}.
Finally, from \eqref{eucZ} and \eqref{eucl-AdSCFT} we obtain for the gauge/gravity prescription for a wormhole
at finite temperature
\be
\label{thermalAdSCFT2}
\mathsf{Tr} \,\left[\, e^{-\beta H}\, \,e^{-\int d {\bf y}\,\phi_0^+({\bf y}) {\cal O^+({\bf y})} -
\int d {\bf y}\,\phi_0^-({\bf y}){\cal O}^-({\bf y})} \,\right]\sim \, e^{
-\frac12\int d\mathbf{y}\,\, d\mathbf{y}'\phi_0^i(\mathbf{y})\,\tilde\Delta_{ij}(\mathbf{y},\mathbf{y}')\,\phi_0^j(\mathbf{y}')}\,.
\ee
This is the key formula for our argumentation because if one now assumes that the degrees of freedom $\Psi_+,\Psi_-$
are decoupled, this is
\be
\label{decoupled}
H[\Psi_+ ,\Psi_-]= H_+[\Psi_+] + H_-[\Psi_-]\,,
\ee
then $\rho_{_\beta}= e^{-\beta H_+}\,e^{-\beta H_-}$,
and the lhs of \eqref{thermalAdSCFT2} factorizes into a product
of two quantities: one depending on $\phi_0^+$ and one depending on
$\phi_0^-$\footnote{Note that in the absence of interaction  $[{\cal O}_+,{\cal O}_-]=0$ no matter if the
operators insertion points are spacelike/timelike separated.}.
However the gravity computation does not factorize because of the non vanishing $\tilde{\Delta}_{\pm\mp}$ terms.
We interpret this result as manifesting the existence of a non-trivial coupling $H_{int}$ between
the two dual degrees of freedom $\Psi_\pm$: the field theory dual to the
wormhole geometry contains  a non-trivial coupling term between the two boundary degrees of freedom $\Psi_+$ and $\Psi_-$.

~

\subsection*{Highlights and Applications}

The outcome of the above observations is that the wormhole geometry
encodes the description of a dual field theory with two copies of
fundamental fields in interaction. Moreover, the quantum state
described by the wormhole is entangled. In particular, the Euclidean
continuation can be seen as prescription to \emph{separate},  in
the dual field theory, entanglement effects from possible
interaction terms $H_{int}$. A non vanishing Euclidean two point
function between operators located at different boundaries must be
interpreted as originated from an interaction term $H_{int}$, rather
than an entanglement effect. In the following we confront this point
of view with two other relevant geometries
appearing in the literature: the AdS$_{1+1}$
geometry and  the eternal AdS black hole \cite{Eternal}.

~

\noindent {\bf AdS$_{1+1}$ geometry:}
The treatment for the AdS$_{1+1}$ background \eqref{ads2} is entirely analogous
to the one performed for the wormhole case. The Euclidean section of the
global metric \eqref{ads2} has two boundaries upon compactifying
the time direction, and the Euclidean
correlation functions can be explicitly obtained from the formulas (\ref{ads2++}).
By virtue of this the arguments above aply, and we can conclude that AdS$_{1+1}$
is dual to a conformal quantum mechanics composed of two interacting sectors.

~

\noindent {\bf Eternal AdS Black Hole:} The crucial difference between the wormhole \eqref{wh} and the
maximally extended AdS-BH gravity solution is well known \cite{Eternal}:
upon performing the Euclidean continuation of the AdS-BH solution, the existence of a horizon
in Lorentzian signature generates a conical
singularity in the Euclidean manifold, that can only be avoided by demanding a precise periodicity
in Euclidean time.
The resulting Riemannian geometry has inevitably only one asymptotic boundary, and therefore requires
imposing only one asymptotic boundary data $\phi_0$, this indicates the existence of a
single set of degrees of freedom $\Psi$ and a unique Bulk-Boundary  propagator $\tilde\Delta$.
The system  in equilibrium with a thermal bath of fixed temperature (determined by the BH mass)
has a generating function that reads
\be
\label{thermalBH}
\mathsf{Tr} \,\left[e^{-\beta H}  e^{-\int d{\bf y}\, \phi_0({\bf y})\, {\cal O}({\bf y})}\right] \sim \,
e^{-\frac{1}{2}\int d{\bf y}  d{\bf y}'  \,\phi_0 ({\bf y}) \,\tilde{\Delta} ({\bf y},{\bf y}')\, \phi_0 ({\bf y}')  } ,
\ee
The real time (maximally extended BH) solution was analyzed and interpreted in the AdS/CFT context in \cite{Eternal}.
The  second {\it causally disconnected} boundary, present in Lorentzian signature, was understood as supporting the TFD
partners needed for obtaining a thermal
state upon their integration and lead to a doubling of the Hilbert space as $\cal H= H_+\otimes H_-$.
We stress the fact that the second set of degrees of freedom is causally disconnected from the original
zero temperature set; although they appear in Lorentzian signature and give rise to a non-trivial $\Delta_{ij}$, they
are ficticious from a physical point of view. The causal disconnection between the boundaries relates to
the two point function $\Delta_{\pm\mp}$ never becoming singular.

\section{Discussion}
\label{discussion}

We have reviewed the GKPW prescription in the Lorentzian context relating the ambiguity in adding
normalizable modes to \eqref{phi} to the integration contour $\cal C$ required to bypass the integrand
singularities in \eqref{deltatt},\eqref{w++} arising from the existence
of normalizable modes. To compute Lorentzian quantities one needs
to fix a reference contour ${\cal C}_{\sf ref}$  and the two sensible choice are retarded or Feynman. These
choices relate on the QFT side to being computing either response or correlation functions.
When choosing the Feynman path as reference contour, any given contour $\cal C$ differs from
${\cal C}^{ F}$ by contributions from encircling poles, these encircled poles fix the initial and
final states  that appear in the correlation functions (see \eqref{2ptf}).

In section \ref{gkpww} we extended the GKPW prescription to spacetimes
with more than one asymptotic timelike boundaries, in particular we studied the
simplest two boundaries case. We proposed to write the bulk field in terms of the
two independent asymptotic boundary values and  as a toy example we applied the construction
to AdS$_{1+1}$ reobtaining previous results, we afterwards  computed the two
point correlation functions for the Einstein-Gauss-Bonnet wormhole \eqref{wh}.
At this point we must emphasize that the principal difference of
our method with the one performed in \cite{balasu}, for an
AdS$_3$ orbifold, consists in that
we considered the boundary values $\phi_0^\pm$ of the bulk scalar
field as independent quantities, moreover, we explicitly showed
the possibility of constructing
two boundary-bulk propagators ${\sf K}^\pm$
(their boundary conditions were given in \eqref{condition2boundaries}).
The construction performed in \cite{balasu} showed a relation between
the boundary values $\phi_0^\pm$ and this was understood as indicating that the $\pm$ sources
for the dual field theory are turned on in a correlated way.
A question remained as whether the two sets of data are
independent or redundant in the dual formulation, on the other hand
another question is the origin of the non-zero result for operators located
at different boundaries
($\Delta_{\pm\mp}$) this could either be due to entanglement or interactions or
both.

In section \ref{holentanglement} we applied the ideas on holographic
entanglement entropy developed in \cite{takaya} to the wormhole
geometry.  The non vanishing of ${\sf S}_{_{\tilde \Sigma}}$
obtained for the degrees of freedom living on opposite boundaries
suggests that the wormhole should be understood as representing an
entangled state in $\cal H=\cal H_+\otimes\,\cal H_-$. On the other
hand,  the  causal connection between the boundaries  suggest that a
coupling might exist as well. To attack this issue we consider
putting the wormhole system in contact with a thermal bath, upon
Euclideanization, the resulting geometry still has two boundaries
connected through the bulk, this indicated, to avoid a
contradiction, that the dual QFT consists of two copies ${\cal H}_{\pm}$
in interaction. Summarizing, the number of disconnected boundaries
of the Euclidean section determines  the amount of physical degrees
of freedom.

We would like to emphasize finally the implications of this approach
on quantum gravity, which could be seen as one of our main
motivations. This subject has been discussed in different contexts
in the last years and referred to as Emergent Spacetime
\cite{seiberg}. In this sense, we showed how important topological
and causal properties (connectivity) of the space time geometry are
encoded in the QFT action, and that part of this information is not
only in the ground state but in its interacting structure. We hope
this conclusion might contribute to the construction of rules
towards a geometry engineering.

 We should mention that in the presence of interactions one needs to
address the way the points on opposite boundaries are identified. A
first approach to this problem is to identify the points $\bf y,y'$
in configuration space at which $\Delta_{\pm\mp}$ diverges and study
the consistency of identifying them, this is currently under
investigation and will be reported elsewhere.

\section*{Acknowledgments}
We thank D. Correa, N. Grandi, A. Lugo and J.M. Maldacena for useful
discussions and correspondence. This work was partially supported by
ANPCyT PICT 2007-0849 and CONICET PIP 2010-0396.

\end{document}